\documentclass[journal,twocolumn]{IEEEtran}

\usepackage{amsmath}
\usepackage{graphicx}
\usepackage{cite}
\usepackage{multirow}
\usepackage{booktabs}
\usepackage{amssymb}
\usepackage{enumitem}
\usepackage{url}
\usepackage{hyperref}
\usepackage{orcidlink}
\usepackage{xcolor}

\begin{document}

\title{Universal Robust Speech Adaptation for Cross-Domain Speech Recognition and Enhancement}

\author{
Chien-Chun~Wang\,\orcidlink{0009-0000-1392-0058},~\IEEEmembership{Student Member,~IEEE,}
Hung-Shin~Lee\,\orcidlink{0000-0001-7044-9434}, \\
Hsin-Min~Wang\,\orcidlink{0000-0003-3599-5071},~\IEEEmembership{Senior Member,~IEEE,}
and~Berlin~Chen\,\orcidlink{0000-0003-0693-8932},~\IEEEmembership{Member,~IEEE}
\thanks{Manuscript received June 23, 2025; revised October 7, 2025, December 24, 2025, and January 27, 2026; accepted February 3, 2026. Date of publication current date; date of current version current date. This work was supported in part by the National Science and Technology Council of Taiwan under Grant NSTC 112-2221-E-001-009-MY3. The
associate editor coordinating the review of this manuscript and approving it for
publication was Dr. Nobutaka Ito. \textit{(Corresponding author: Hung-Shin Lee.)}}
\thanks{Chien-Chun Wang and Berlin Chen are with the Department of Computer Science and Information Engineering, National Taiwan Normal University, Taipei 11677, Taiwan (e-mail: jethrowang0531@gmail.com; berlin@ntnu.edu.tw).}
\thanks{Hung-Shin Lee is with United Link Co., Ltd., Taipei 11493, Taiwan (e-mail: hungshinlee@gmail.com).}
\thanks{Hsin-Min Wang is with the Institute of Information Science, Academia Sinica, Taipei 11529, Taiwan (e-mail: whm@iis.sinica.edu.tw).}
}

% \markboth{IEEE/ACM Transactions on Audio, Speech, and Language Processing, Vol. ??, 2026}
% {Wang \MakeLowercase{\textit{et al.}}: Universal Robust Speech Adaptation for Cross-Domain Speech Recognition and Enhancement}

\maketitle

\begin{abstract}

Pre-trained models for automatic speech recognition (ASR) and speech enhancement (SE) have exhibited remarkable capabilities under matched noise and channel conditions.
However, these models often suffer from severe performance degradation when confronted with domain shifts, particularly in the presence of unseen noise and channel distortions.
In view of this, we in this paper present URSA-GAN, a unified and domain-aware generative framework specifically designed to mitigate mismatches in both noise and channel conditions.
URSA-GAN leverages a dual-embedding architecture that consists of a noise encoder and a channel encoder, each pre-trained with limited in-domain data to capture domain-relevant representations.
These embeddings condition a GAN-based speech generator, facilitating the synthesis of speech that is acoustically aligned with the target domain while preserving phonetic content.
To enhance generalization further, we propose dynamic stochastic perturbation, a novel regularization technique that introduces controlled variability into the embeddings during generation, promoting robustness to unseen domains.
Empirical results demonstrate that URSA-GAN effectively reduces character error rates in ASR and improves perceptual metrics in SE across diverse noisy and mismatched channel scenarios.
Notably, evaluations on compound test conditions with both channel and noise degradations confirm the generalization ability of URSA-GAN, yielding relative improvements of 16.16\% in ASR performance and 15.58\% in SE metrics.

\end{abstract}


% \IEEEpeerreviewmaketitle

\section{Introduction}

\begin{figure}[t]
\centering
\includegraphics[width=1.0\linewidth]{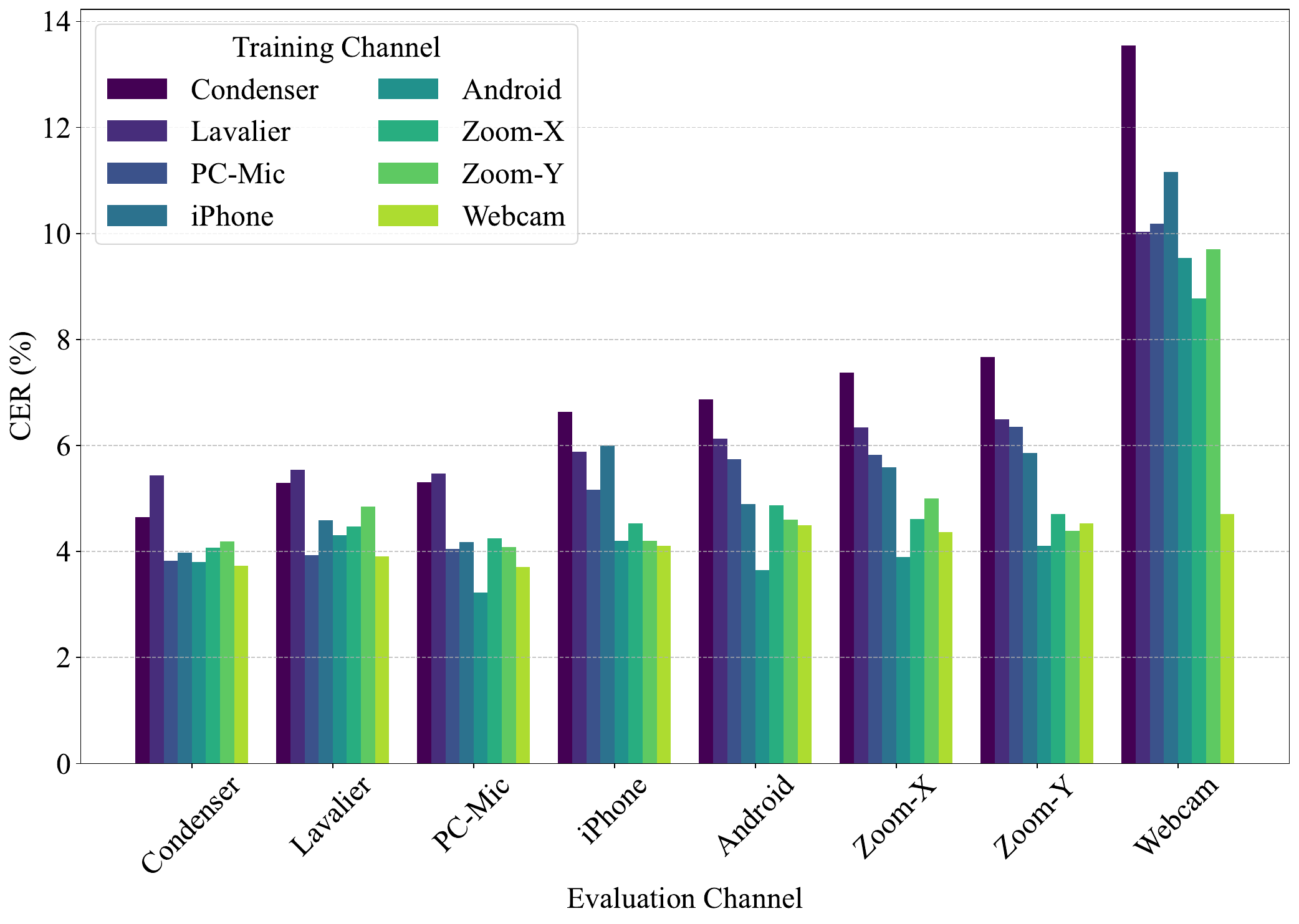}
\vspace{-20pt}
\caption{
Character error rates (CERs) of ASR models on the HAT corpus.
Each group of bars represents evaluation on audio from a specific recording device, with individual bars showing CERs for models trained on different devices.
Performance generally degrades under device mismatch, underscoring the impact of channel variation and the challenge of cross-device generalization.
}
\label{fig:channels_eval}
\vspace{-10pt}
\end{figure}

\IEEEPARstart{S}{ystems} encompassing automatic speech recognition (ASR) and speech enhancement (SE) have made great strides in recent years due to the widespread adoption of deep learning techniques.
For instance, models based on convolutional neural networks (CNNs) \cite{park2016,tan2019,pan2020}, recurrent neural networks (RNNs) \cite{hu2018,zhao2018,park2020,zeyer2021}, Transformers \cite{gulati2020,baevski2020,chan2021,haidar2021,xu2021,de2022,kim2022,radford2023}, and generative adversarial networks (GANs) \cite{pascual2017,fu2019,xiang2020,cao2022,zadorozhnyy2022}, to name a few, have achieved remarkable performance across various speech-related use cases.
However, these systems often experience considerable performance degradation when exposed to mismatched conditions, particularly in the presence of unseen noise types or acoustic channel variations, as illustrated in Fig. \ref{fig:channels_eval}.
These phenomena highlight a critical limitation in practical applications, where environmental noise and recording equipment frequently differ from those conditions observed during model training.
The issue of domain mismatch, stemming from such variability, significantly compromises the robustness and generalization capabilities of contemporary ASR and SE systems.
To address this challenge, a robust and practical modeling framework for ASR and SE is required that can effectively adapt to both noise and channel distortions.

Conventional domain adaptation approaches \cite{wang2015,long2016,tzeng2017,hsu2018,mun2019,shu2019,li2020,hou2021,wang2022} have aimed to address domain mismatches through strategies such as adversarial training, feature-space transformation, and data augmentation.
While these approaches have demonstrated promising results, they often rely on a significant amount of labeled target-domain data or entail complex training procedures that impede their scalability in practical applications.
To overcome these limitations, data simulation has recently emerged as a viable alternative solution for domain adaptation in both ASR \cite{hu2018,chen2022} and SE \cite{chen2023}.
These methods generate synthetic speech that approximates the acoustic characteristics of the target domain, facilitating model adaptation without the need for extensive real-world recordings.
Nonetheless, most current simulation techniques primarily focus on capturing broad domain properties and often neglect fine-grained, utterance-level variations that are crucial for robust generalization.
Moreover, while our previous studies \cite{wang2024,wang2025} have respectively focused on robust adaptation to either noisy environments or channel mismatches, they address these challenges independently and do not consider their combined effect.
Similarly, most existing methods in the literature tend to treat noise and channel distortion in isolation, lacking a unified framework that handles both simultaneously.

To bridge the gap and build upon our earlier studies \cite{wang2024,wang2025}, we propose the \textbf{U}niversal \textbf{R}obust \textbf{S}peech \textbf{A}daptation \textbf{G}enerative \textbf{A}dversarial \textbf{N}etwork (URSA-GAN), a novel framework that jointly models both environmental noise and channel distortions for unified domain adaptation in ASR and SE tasks.
URSA-GAN adopts a two-stage training pipeline.
In the first stage, two dedicated encoders are trained to separately extract noise and channel embeddings from unlabeled speech in the target domain.
These embeddings are designed to capture instance-level distortions and are disentangled to reflect distinct acoustic properties.
In the second stage, these embeddings guide a GAN-based generator to synthesize seemingly realistic speech that reflects the target-domain noise and channel characteristics while preserving phonetic content.
A discriminator ensures the realism of the generated speech by enforcing distributional alignment with actual target-domain recordings.
Unlike other methods \cite{chen2022,chen2023}, which employ global transformations to approximate target conditions, URSA-GAN learns fine-grained, utterance-specific embeddings that facilitate more precise and robust simulations.
Furthermore, to enhance generalization to unseen environments, we introduce dynamic stochastic perturbation, which injects controlled variability into the embeddings during the generation process.
This mechanism encourages the model to produce a smoother and more comprehensive representation of target-domain variability, thereby improving its adaptability in previously unencountered settings.

Extensive evaluations are conducted across diverse benchmarks covering both channel and noise mismatches.
For ASR, we assess performance under cross-domain channel conditions, while for SE, the noisy speech scenarios are considered with varying acoustic environments.
A challenging hybrid evaluation set is also constructed to combine both noise and channel degradations.
Across all cases, URSA-GAN consistently demonstrates robustness and generalization, validating its effectiveness in complex real-world scenarios.
In summary, the key contributions of this study are outlined as follows:
\begin{enumerate}[label=\arabic*),leftmargin=14pt,itemsep=0ex,topsep=4pt]
\item \textbf{Unified Noise-Channel Adaptation:}
We propose a novel framework that jointly adapts to environmental noise and channel distortions, leveraging instance-level embeddings to simulate realistic target-domain conditions. 
The unified framework builds upon our previous studies \cite{wang2024,wang2025} to enhance the robustness of both ASR and SE systems.
\item \textbf{Efficient and Generalizable Learning:}
Our framework achieves strong performance with minimal unlabeled target-domain data, thanks to the data-efficient learning and dynamic stochastic perturbation mechanisms that improve generalization to unseen environments.
\item \textbf{Broad and Rigorous Evaluation:}
Our framework is comprehensively evaluated across multiple datasets and tasks under both isolated and combined distortion settings, demonstrating its scalability and versatility.
\end{enumerate}

\section{Preliminaries}

Automatic speech recognition and speech enhancement are central tasks in speech processing, aiming to transcribe spoken language and improve speech quality under adverse conditions.
Although recent deep learning approaches have led to notable improvements on ASR and SE, their real-world performance remains limited by mismatches in noise and recording channels.
Domain adaptation techniques have been proposed to address these issues, typically focusing on either noise or channel mismatch in isolation.
However, practical applications often involve both simultaneously, which significantly challenges model robustness.
This study is motivated by the need for a unified framework that can jointly address these mismatches to improve ASR and SE performance in realistic environments.

\subsection{Automatic Speech Recognition}

End-to-end neural architectures have established themselves as the prevailing paradigm in contemporary ASR systems, yielding enhanced performance and streamlined workflows.
Among these models, Whisper \cite{radford2023} stands out as a particularly effective solution, trained on extensive multilingual and noisy datasets.
Its exceptional generalization capabilities and availability to the research community have rendered it a preferred choice in studies targeting robustness and data scarcity.
In this study, we employ Whisper due to its strong empirical performance and inherent robustness, which facilitates a more rigorous evaluation of domain adaptation strategies.
By mitigating the effects of recognition model constraints, we can more directly evaluate the efficacy of our proposed adaptation techniques.
While other recent ASR models, such as OWSM, OWSM-CTC, Canary, and OWLS \cite{peng2023,peng2024,puvvada2024,chen2024}, present promising alternatives, Whisper continues to serve as a widely recognized and representative benchmark, especially in contexts involving multilingual and noisy inputs.

\begin{figure*}[ht]
\centering
\includegraphics[width=1.0\linewidth]{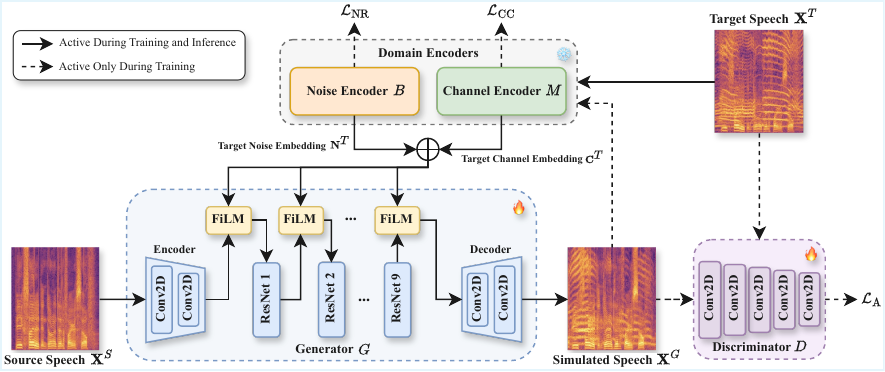}
\vspace{-15pt}
\caption{
The architecture of our proposed framework, URSA-GAN.
Solid lines represent the forward data flow during both training and inference phases.
The dashed arrows indicate that during the training phase, simulated speech $\mathbf{X}^G$ is used together with target speech $\mathbf{X}^T$ to 1) train the discriminator $D$, and 2) contribute to noise reconstruction and channel consistency. The $\bigoplus$ operator denotes element-wise tensor addition.
}
\label{fig:URSA-GAN}
\vspace{-10pt}
\end{figure*}

\subsection{Speech Enhancement}

Mainstream SE models are primarily developed using frequency-domain approaches.
In these models, noisy spectral features serve as the input, while the learning target is either the clean spectral features or a corresponding mask \cite{yin2020}.
DEMUCS \cite{defossez2020} is one of the leading frequency-domain models, leveraging both convolutional and recurrent layers to perform high-fidelity waveform restoration.
Its strong denoising performance and ability to generalize across diverse conditions make it a practical and reliable choice for real-world applications.
In our framework, DEMUCS is employed to ensure that the enhancement module does not become a limiting factor.
This allows us to isolate and examine the effectiveness of domain adaptation.
Although other SE models, such as FullSubNet+, MP-SENet, CleanUNet 2, and SEMamba \cite{chen2022b,lu2023,kong2023,chao2024}, perform well under controlled environments, DEMUCS offers a favorable combination of quality, generalization, and efficiency, which makes it suitable for real-world use cases.

\subsection{Domain Adaptation: UNA-GAN}

Unsupervised noise adaptation has attracted growing attention for SE when target-domain clean references are unavailable.
UNA-GAN \cite{chen2023} addresses this by learning a clean-to-noisy transformation, converting source-domain clean speech into target-style noisy speech.
Unlike feature alignment or domain-classifier approaches \cite{liao2019,hou2021,yang2025}, UNA-GAN focuses on noise modeling, enabling large-scale parallel data simulation for adaptation.
Its framework consists of a GAN-based simulator and an SE model.
The generator incorporates target-domain noise characteristics into clean inputs using convolutional, ResNet, and self-attention blocks, while a discriminator enforces realistic noise distributions.
A contrastive loss preserves speech content during simulation.
With only a few minutes of unpaired target data, this design achieves high data efficiency and supports unpaired training.
The simulated clean-noisy pairs are then used to finetune a TCN-based SE model \cite{luo2019}, effectively mitigating domain mismatch.
Experiments in \cite{chen2023} show substantial robustness gains under low-SNR and large mismatch conditions, establishing UNA-GAN as a strong baseline for data simulation-based adaptation.

\section{Proposed Methodology}

\subsection{Framework Overview}

Fig. \ref{fig:URSA-GAN} schematically depicts the overall architecture of the proposed URSA-GAN, which is composed of four main components, including a generator $G$, a discriminator $D$, a noise encoder $B$, and a channel encoder $M$.
For clarity and ease of reference, the key notations used throughout this paper are summarized in Table \ref{tab:notation}.
The primary objective of this framework is to simulate target-domain speech $\mathbf{X}^T$ by transferring noise and channel characteristics onto source-domain input $\mathbf{X}^S$, thereby enhancing the robustness of downstream models.
Given a target-domain spectrogram, the noise encoder extracts a noise embedding $\mathbf{N}^T$ that captures environmental interference, while the channel encoder derives a channel embedding $\mathbf{C}^T$ that models transmission-related distortions. As such, these embeddings encapsulate the fundamental acoustic characteristics of the target domain $T$.
The generator receives a clean source-domain spectrogram along with the target-domain embeddings and produces a simulated spectrogram $\mathbf{X}^G$.
The generated spectrogram preserves the linguistic content of $\mathbf{X}^S$ while incorporating the noise and channel characteristics from $\mathbf{X}^T$, effectively creating domain-adapted speech for training purposes.
The discriminator is trained to distinguish between real target-domain spectrograms and generated ones, thereby encouraging the generator to produce more realistic outputs through adversarial learning.
Most notably, our framework enables instance-level, label-free domain simulation by leveraging a minimal amount of target-domain data, thereby offering a scalable solution for effective speech adaptation in both ASR and SE tasks.

\begin{table}[t]
\small
\caption{Summary of notations used in this paper.}
\vspace{-5pt}
\label{tab:notation}
\centering
\setlength{\tabcolsep}{7.5pt}
\begin{tabular}{l|l}
\toprule 
\textbf{Symbol} & \textbf{Description} \\
\toprule
$\mathbf{X}^S$ & Source-domain clean speech spectrogram \\
$\mathbf{X}^T$ & Target-domain speech spectrogram \\
$\mathbf{X}^G$ & Simulated (generated) target-domain spectrogram \\
$\mathbf{N}^T$ & Target noise embedding extracted by encoder $B$ \\
$\mathbf{C}^T$ & Target channel embedding extracted by encoder $M$ \\
$G$ & Generator network \\
$D$ & Discriminator network \\
$B$ & Noise encoder (based on BEATs) \\
$M$ & Channel encoder (based on MFA-Conformer) \\
$\lambda_{\text{NR}}$ & Weighting factor for noise reconstruction loss \\
$\lambda_{\text{CC}}$ & Weighting factor for channel consistency loss \\
\bottomrule
\end{tabular}
\vspace{-10pt}
\end{table}

\subsection{Generator and Discriminator}

The generator $G$ is designed to transform a source-domain spectrogram $\mathbf{X}^S$ into a simulated target-domain spectrogram $\mathbf{X}^G$.
The architecture of the generator follows an encoder-decoder structure with residual connections to facilitate efficient transformation while preserving crucial spectral characteristics.
Initially, the input spectrogram is processed through two 2D downsampling convolutional layers (kernel size: $3\times3$, stride: $2\times2$), which reduce the spatial dimensions while effectively extracting low-level spectral features.
Following these, a series of nine residual blocks come into play, where each block comprises two convolutional layers (kernel size: $3\times3$, stride: $1\times1$) equipped with instance normalization and ReLU activation.
This design choice ensures stable training and the preservation of high-resolution feature representations.
A dropout layer is incorporated within each residual block to prevent overfitting and enhance generalization.
Finally, two transposed convolutional layers (kernel size: $3\times3$, stride: $2\times2$) are employed to upsample the learned representations, thereby restoring the original spectrogram dimensions while ensuring the generated spectrogram exhibits realistic characteristics.
In addition to the primary spectral transformation, the generator incorporates domain-specific noise $\mathbf{N}^T$ and channel information $\mathbf{C}^T$ during training.
This strategy enables the simulation of real-world variations in the target spectrograms $\mathbf{X}^T$.
The combination of these auxiliary features ensures that the generated spectrogram retains environmental robustness and domain adaptability.

The discriminator $D$ is pivotal in differentiating between authentic target spectrograms and synthesized spectrograms.
It consists of five 2D convolutional layers (kernel size: $4\times4$) with Leaky ReLU activation functions.
The stride is set to $2\times2$ for the first three layers and $1\times1$ for the last two, progressively increasing the receptive field while preserving fine-grained spectral details.
Batch normalization is omitted to avoid mode collapse, ensuring that the discriminator learns meaningful distinctions between real and generated spectrograms.
The final output is a scalar probability score indicating the authenticity of the input spectrogram.

The adversarial loss function used during training is defined as follows:
\begin{equation}
\begin{split}
\mathcal{L}&_{\text{A}}(G,D,\mathbf{X}^T,\mathbf{X}^S,\mathbf{N}^T,\mathbf{C}^T) = \mathbb{E}_{\mathbf{x} \sim \mathbf{X}^T} \left[ \log D(\mathbf{x}) \right] \\
&+ \mathbb{E}_{\mathbf{x} \sim \mathbf{X}^S,\mathbf{n} \sim \mathbf{N}^T,\mathbf{c} \sim \mathbf{C}^T} \left[ \log (1-D(G(\mathbf{x},\mathbf{n},\mathbf{c}))) \right].
\end{split}
\end{equation}
This loss encourages the generator to produce spectrograms that closely resemble target spectrograms, while the discriminator learns to identify subtle spectral differences.
To stabilize training and prevent vanishing gradients, we employ spectral normalization in the discriminator and instance normalization in the generator.
Additionally, gradient penalty is introduced to further regularize the discriminator, ensuring a well-balanced adversarial training process.
This ongoing interaction between the generator and the discriminator results in progressively refined spectrogram synthesis, thereby improving the quality of domain adaptation for speech processing tasks.

\subsection{Domain Encoders}

To explicitly capture domain-specific acoustic variations in target environments, we introduce an encoder module composed of two specialized domain encoders, the noise encoder $B$ and the channel encoder $M$.
These encoders aim to extract meaningful embeddings that represent noise and channel characteristics, respectively, which are subsequently used to guide the generative process.

The noise encoder is designed to capture detailed noise attributes from the target spectrogram $\mathbf{X}^T$.
Inspired by recent advances in noise-aware modeling \cite{li2021,wang2023,hu2024}, the encoder extracts noise embeddings $\mathbf{N}^T$ to provide the generator $G$ with a precise representation of the environmental noise.
To be specific, we employ BEATs \cite{chen2023a}, a pre-trained audio model known for robust acoustic feature extraction, as the backbone of our noise encoder.
BEATs is particularly well-suited for this role due to its acoustic event–centric pre-training strategy, which fundamentally differs from self-supervised learning (SSL) models primarily optimized for phonetic or linguistic content.
Unlike speech-focused SSL encoders that emphasize phoneme discrimination or contextual linguistic representations, BEATs is trained to capture general-purpose acoustic structures across diverse sound events.
This design enables BEATs to retain rich representations of background noise and non-speech sounds, making it especially effective for modeling domain-specific acoustic variations rather than speaker- or content-related attributes.
Such properties are critical in cross-domain adaptation scenarios, where robustness to non-linguistic acoustic factors plays a central role.

To enhance its noise discrimination capability, we further fine-tune the BEATs-based noise encoder beyond its original pre-training.
While BEATs is effective at capturing general acoustic structures, target-domain adaptation is necessary to model domain-specific noise characteristics that are not sufficiently represented in large-scale pre-training data.
To mitigate potential catastrophic forgetting during fine-tuning, we adopt a controlled and progressive adaptation strategy.
Specifically, BEATs is fine-tuned with a relatively small learning rate while keeping its architectural core unchanged, ensuring that the general acoustic knowledge acquired during large-scale pre-training is preserved.
Within this constrained setting, we employ a two-stage fine-tuning procedure to balance generalization and specialization.
In the first stage, the model is refined on the ESC-50 dataset, which contains diverse environmental sound classes, reinforcing its ability to discriminate a broad range of common acoustic noise patterns and stabilizing the noise embedding space.
In the second stage, BEATs is further adapted by treating each utterance in the target-domain data used during GAN training as a distinct noise class.
Rather than reshaping the representation space, this stage focuses on fine-grained specialization, enabling BEATs to capture domain-specific noise characteristics without overwriting the general acoustic structures learned in earlier stages.
The extracted noise embeddings are integrated into the generator together with a source-domain spectrogram.
The generator produces a synthesized spectrogram that incorporates the noise characteristics while preserving the underlying phonetic content.

To enforce faithful noise representation, we introduce a noise reconstruction loss defined by
\begin{equation}
\begin{split}
&\mathcal{L}_{\text{NR}}(G,\mathbf{X}^S,\mathbf{N}^T,\mathbf{C}^T) \\ 
&= \mathbb{E}_{\mathbf{x} \sim \mathbf{X}^S, \mathbf{n} \sim \mathbf{N}^T, \mathbf{c} \sim \mathbf{C}^T} \left[ \left\| \mathbf{n} - B(G(\mathbf{x},\mathbf{n},\mathbf{c})) \right\|_1 \right].
\end{split}
\end{equation}
This loss reinforces that the noise embeddings extracted from the generated speech closely match the original target-domain noise embeddings, thereby encouraging the generator to maintain accurate noise details in the synthesis.

Complementing the noise encoder, our channel encoder focuses on capturing channel-related distortions caused by microphone variability, transmission interference, and others.
Unlike conventional methods that use raw spectrograms as channel proxies, we specifically leverage MFA-Conformer \cite{zhang2022}, pre-trained on the HAT corpus \cite{liao2023}, where multiple speakers utter identical content recorded simultaneously on different microphones.
This training setup allows the channel encoder to learn channel-specific embeddings $\mathbf{C}^T$ that are invariant to speaker identity and phonetic content.
The MFA-Conformer architecture effectively models both local and global acoustic dependencies, aggregating features across multiple temporal scales to enhance sensitivity to subtle channel variations.
By intentionally omitting the source and target channels from the training data employed in our primary experiments, the channel encoder is designed to generalize effectively to unexplored channel conditions.

To promote accurate channel adaptation for use in speech synthesis, we define a channel consistency loss:
\begin{equation}
\begin{split}
&\mathcal{L}_{\text{CC}}(G,\mathbf{X}^S,\mathbf{N}^T,\mathbf{C}^T) \\ 
&= \mathbb{E}_{\mathbf{x} \sim \mathbf{X}^S, \mathbf{n} \sim \mathbf{N}^T, \mathbf{c} \sim \mathbf{C}^T} \left[ \left\| \mathbf{c} - M(G(\mathbf{x},\mathbf{n},\mathbf{c})) \right\|_1 \right].
\end{split}
\end{equation}
This loss constrains the generated spectrograms to retain channel characteristics by minimizing the difference between channel embeddings extracted from synthesized and target-domain speech.

Together, the noise and channel encoders yield fine-grained, disentangled acoustic representations, allowing our model to effectively simulate target-domain speech under realistic noise and channel distortions.
The integration of these domain encoders with the generator through corresponding reconstruction losses enable the synthesized data to faithfully capture both environmental noise and channel conditions, thus enhancing robustness in downstream ASR and SE tasks.

\subsection{Feature Fusion}

To effectively incorporate noise $\mathbf{N}^T$ and channel embeddings $\mathbf{C}^T$ into the generator module $G$, we adopt feature-wise linear modulation (FiLM) \cite{perez2018} as our feature fusion mechanism.
FiLM enables the model to conditionally modulate intermediate feature representations by applying learned affine transformations, which are crucial for adapting the generator to different noise and channel conditions.

At the outset, the noise and channel embeddings are summed element-wise and then processed through separate linear transformations to derive the weight $\mathbf{W}$ and bias $\mathbf{b}$:
\begin{equation}
\mathbf{W} = \mathrm{Linear}(\mathbf{N}^T + \mathbf{C}^T), \quad \mathbf{b} = \mathrm{Linear}(\mathbf{N}^T + \mathbf{C}^T),
\end{equation}
where $\mathrm{Linear}(\cdot)$ denotes a linear transformation.
These transformations allow the model to learn a set of adaptive scaling and shifting parameters based on the given noise and channel conditions.
Subsequently, the output feature $\mathbf{F}$ from a specific layer within the generator is modulated using these parameters as follows:
\begin{equation}
\mathbf{F}' = \mathbf{W} \times \mathbf{F} + \mathbf{b}.
\end{equation}
Here, the feature-wise multiplication with $\mathbf{W}$ dynamically scales each feature channel, while the addition of $\mathbf{b}$ introduces a shift, ensuring that the generator can adjust its feature representation accordingly.

The FiLM-based modulation is applied not only at the encoder output but consistently across all nine ResNet blocks in the generator.
This design choice is motivated by the observation that noise and channel mismatches affect multiple levels of speech representations, including low-level spectral structures, mid-level acoustic patterns, and high-level abstract features.
Restricting the modulation to a single layer would limit the ability of the generator to compensate for such cross-level distortions.
By performing FiLM conditioning at every residual block, the generator maintains continuous awareness of domain-specific acoustic factors throughout the entire transformation pipeline.
Importantly, this multi-layer FiLM strategy introduces only marginal computational overhead.
FiLM performs channel-wise affine transformations whose complexity scales linearly with the number of feature channels, resulting in negligible cost compared to the convolutional and residual operations within each ResNet block.
Therefore, applying FiLM across all generator layers does not significantly increase training or inference complexity while providing effective and fine-grained domain adaptation capability.
Notably, each FiLM block uses independently learned parameters rather than sharing a single set across blocks, allowing each layer to adapt uniquely to the noise and channel embeddings.
By incorporating FiLM across multiple layers with independent parameters, the generator remains aware of noise and channel variations throughout the feature transformation process, ultimately producing speech signals that are more robust against environmental distortions and improving performance for downstream tasks.

\subsection{Patch-wise Contrastive Learning}

To preserve linguistic consistency between the generated speech $\mathbf{X}^G$ and the original clean speech $\mathbf{X}^S$, we adopt patch-wise contrastive learning (PCL) \cite{park2020a}.
The goal is to maximize the mutual information between source and simulated spectrograms, guiding the generator $G$ to retain critical phonetic and structural features while suppressing distortions.
Unlike frame-level supervision, PCL operates on patches, where each patch denotes a short block of consecutive frames extracted from a feature map.
This design captures both local temporal continuity and spectral structure, enabling fine-grained alignment between source and simulated speech.

\begin{figure}[t]
\centering
\includegraphics[width=1.0\linewidth]{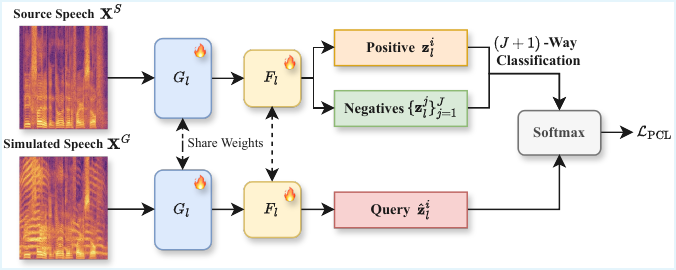}
\vspace{-20pt}
\caption{
Illustration of the patch-wise contrastive learning process.
Solid arrows denote the flow of feature extraction and projection, while dashed arrows indicate shared weights between the projection heads.
The process involves feature extraction using the generator $G$, patch sampling, projection into the embedding space through the projection head $F$, and cross-entropy loss calculation using the softmax activation function.}
\label{fig:pcl}
\vspace{-10pt}
\end{figure}

As illustrated in Fig.~\ref{fig:pcl}, the PCL procedure proceeds as follows.
The generator extracts hierarchical feature maps $G_l(\mathbf{X})$ from both source and simulated speech across $L=4$ layers, including two down-sampling convolutional layers and two residual blocks, which together capture both low-level spectral patterns and high-level semantic content.
From each feature map, we randomly sample $I=256$ query patches from the simulated representation $G_l(\mathbf{X}^G)$.
The corresponding patches from the source feature map $G_l(\mathbf{X}^S)$ are regarded as positive samples, while $J$ other mismatched patches are treated as negatives.
All query, positive, and negative patches are projected into a lower-dimensional embedding space through a layer-specific projection head $F_l$, consisting of two linear layers with 256 units and a ReLU activation.
Formally, the patch embedding process is expressed as
\begin{equation}
\begin{aligned}
\hat{\mathbf{z}}_l^i &= F_l(G_l(\mathbf{X}^G)^{[i]}), \\
\mathbf{z}_l^i &= F_l(G_l(\mathbf{X}^S)^{[i]}), \\
\mathbf{z}_l^j &= F_l(G_l(\mathbf{X}^S)^{[j]}),
\end{aligned}
\end{equation}
where $(\cdot)^{[i]}$ denotes the $i$-th consecutive frame patch sampled from the feature map at layer $l$.
This construction yields a set of query-positive pairs $(\hat{\mathbf{z}}_l^i, \mathbf{z}_l^i)$ and query-negative pairs $(\hat{\mathbf{z}}_l^i, \mathbf{z}_l^j)$.
The training objective is formulated as a $(J+1)$-way classification task at each layer, where the model is required to identify the true positive patch among $J$ negatives.
The cross-entropy loss is defined as
\begin{equation}
\mathcal{L}_{\text{PCL}}(G,\mathbf{X}^S) = \sum\limits_{l=1}^{L} \sum\limits_{i=1}^{I} -\log \frac{e^{\left( \hat{\mathbf{z}}_l^i \cdot \mathbf{z}_l^i / \tau \right)}}{e^{\left( \hat{\mathbf{z}}_l^i \cdot \mathbf{z}_l^i / \tau \right)} + \sum_{j=1}^{J} e^{\left( \hat{\mathbf{z}}_l^i \cdot \mathbf{z}_l^j / \tau \right)}},
\end{equation}
where $\tau$ is a temperature parameter controlling the sharpness of discrimination.
In addition, this loss function is applied not only to the source spectrogram but also to the target-domain spectrogram.
Our dual application regularizes the generator to produce intelligible speech across diverse acoustic conditions, ensuring that outputs remain consistent in linguistic content despite noise and channel variations.
By combining patch-level discriminability with multi-layer supervision, PCL imposes a strong constraint that complements adversarial and reconstruction objectives, leading to more robust and high-fidelity speech synthesis.

\subsection{Training Objective}

During the training phase, the primary objective is to optimize the GAN model by leveraging a well-designed loss function that balances multiple learning criteria.
The overall loss function, denoted as $\mathcal{L}_{\text{Overall}}$, integrates adversarial learning, patch-wise contrastive learning, noise reconstruction, and channel consistency to ensure that the generated spectrograms $\mathbf{X}^G$ retain crucial speech characteristics while effectively modeling both noise and channel distortions.
The formulation of the overall loss is expressed as:
\begin{equation}\label{eq_overall}
\begin{split}
\mathcal{L}_{\text{Overall}} &= \mathcal{L}_{\text{A}}(G,D,\mathbf{X}^T,\mathbf{X}^S,\mathbf{N}^T,\mathbf{C}^T) \\
&+ \mathcal{L}_{\text{PCL}}(G,\mathbf{X}^S) + \mathcal{L}_{\text{PCL}}(G,\mathbf{X}^T) \\
&+ \lambda_{\text{NR}}\mathcal{L}_{\text{NR}}(G,\mathbf{X}^S,\mathbf{N}^T,\mathbf{C}^T) \\
&+ \lambda_{\text{CC}}\mathcal{L}_{\text{CC}}(G,\mathbf{X}^S,\mathbf{N}^T,\mathbf{C}^T),
\end{split}
\end{equation}
where $\lambda_{\text{NR}}$ and $\lambda_{\text{CC}}$ are weighting factors that control the contributions of noise reconstruction loss and channel consistency loss, respectively.
Given the scarcity of target-domain data, we employ a strategy where an equal number of speech samples are randomly selected from the source domain to construct a balanced dataset.
The training process is then conducted using this unpaired dataset, with optimization driven by the overall loss function $\mathcal{L}_{\text{Overall}}$ in Eq. \ref{eq_overall}.
The strategy allows the model to generalize effectively across different domains while maintaining robustness against noise and channel variations.

\subsection{Adaptation Process}

After training, the well-trained generator $G$ serves as a domain converter, mapping the source speech $\mathbf{X}^S$ to the target speech $\mathbf{X}^T$.
This conversion process is facilitated by the fine-tuned noise encoder $B$ and the channel encoder $M$, which work together to simulate speech data $\mathbf{X}^G$ that closely mirrors the characteristics of the target domain $T$.
The generator, in combination with these domain encoders, synthesizes speech by leveraging a rich pool of source-domain speech data and randomly selected target-domain speech from the training phase.
This results in a sufficient amount of paired data consisting of clean speech and simulated target-domain speech, which is crucial for achieving accurate domain adaptation.
The augmented dataset formed in this manner can then be used to fine-tune any downstream ASR and SE models, enabling them to adapt effectively to the target domain.

In addition to domain conversion, dynamic stochastic perturbation is employed during the generation stage to further enhance the adaptability and robustness of the downstream SE models.
The mechanism introduces Gaussian noise with a variable standard deviation to the embeddings extracted from the target-domain spectrograms.
The standard deviation governs the extent of the spread of the disturbance, allowing flexible control over the interference levels.
By incorporating dynamic stochastic perturbation into the generation process, the model can simulate a range of interference conditions, which is crucial for preventing overfitting to specific noise or channel patterns that were encountered during training.
Moreover, the process enhances the resilience of our model to unseen conditions, strengthening its generalization capabilities and guaranteeing that it performs well across various challenging acoustic environments.
Therefore, the proposed mechanism is instrumental in maintaining the adaptability of the downstream SE models when deployed in real-world scenarios where unforeseen noise sources or channel distortions may occur.

\section{Experimental Setups}

\subsection{Datasets}

\subsubsection{Hakka Across Taiwan (HAT)}

The HAT corpus \cite{liao2023} contains speech recorded from eight microphones, capturing diverse acoustic conditions. 
These include an iPhone, an Android phone, a webcam, a professional condenser microphone, a lavalier microphone, a budget PC microphone (PC-mic), and an X-Y stereo microphone (ZOOM-X and ZOOM-Y). 
Each channel has 97,385 training and 4,559 test utterances, totaling 779,080 training and 36,472 test utterances across all channels. 
Recordings from the condenser microphone were designated as the source domain and those from the webcam as the target domain, as this pair exhibits the largest mismatch.
For GAN training, 40 utterances were randomly sampled from each domain to demonstrate the effectiveness of our framework with limited target-domain data.

\subsubsection{Taiwanese Across Taiwan (TAT)}

To further verify that our channel encoder does not unintentionally encode phonetic information, we conducted additional experiments using the TAT corpus \cite{liao2022}.
The corpus is similar to HAT but does not include recordings from webcam and PC-mic.
For this evaluation, we designated recordings from the condenser microphone as the source domain and those from Android devices as the target domain.
In particular, there is no indication that HAT and TAT share the same device types or brands.

\subsubsection{VoiceBank-DEMAND (VBD)}

The VBD dataset \cite{valentini2016} is a widely recognized benchmark for speech enhancement, created by mixing clean speech samples from the VoiceBank corpus \cite{valentini2016} with noise samples from the DEMAND database \cite{thiemann2013}.
VBD encompasses recordings from 30 speakers, with 28 allocated for training and 2 set aside for testing.
The training set (source domain) comprises 11,572 utterances, generated by mixing clean speech with 10 noise types at four signal-to-noise ratio (SNR) levels (0, 5, 10, and 15 dB).
In contrast, the test set (target domain) contains 824 utterances featuring five previously unseen noise types (living room, office space, bus, open area cafeteria, and public square) at four different SNR levels (2.5, 7.5, 12.5, and 17.5 dB).
For GAN training, 40 utterances were randomly sampled from the test set as target-domain data, and subsequently removed from evaluation, leaving 784 utterances for final evaluation.

\subsubsection{HAT-ESC}

The ESC-50 noise dataset \cite{piczak2015} was integrated with the HAT corpus to construct a new benchmark dataset, termed HAT-ESC, designed to capture both noise and channel distortions simultaneously.
ESC-50 consists of 2,000 labeled environmental audio recordings, each 5 seconds long, organized into 50 semantic classes with 40 examples per class.
It is a widely used benchmark for environmental sound classification.
In the construction of HAT-ESC, five specific noise types from the ESC-50 dataset were randomly selected as the target domain.
These include chirping birds, church bells, laughter, pouring water, and a vacuum cleaner.
The remaining noise types were treated as the source domain.
Channel-distorted speech data were combined with these environmental noises using random SNRs of 0, 5, 10, and 15 dB.
This combination allows URSA-GAN to learn more effective noise and channel adaptation strategies, and enhances robustness in real-world ASR and SE tasks.

\begin{table}[t]
\small
\caption{Parameter counts of different components in URSA-GAN.}
\vspace{-5pt}
\label{tab:parameter}
\centering
\setlength{\tabcolsep}{32pt}
\begin{tabular}{lc}
\toprule 
\bf{Component} & \bf{Parameter (M)} \\
\toprule
Generator & 50.7 \\
Discriminator & 2.8 \\
Noise Encoder & 90.0 \\
Channel Encoder & 20.5 \\
Projection Head & 0.6 \\
\bottomrule
\end{tabular}
\vspace{-10pt}
\end{table}

\subsection{Model Configuration and Resource Allocation}

To effectively capture short-time characteristics of noise and channel variations, the input speech spectrograms were segmented into patches of $129$ frequency bins by $128$ time frames.
This allows our domain encoders to extract fine-grained, domain-specific information from both spectral and temporal dimensions.
The GAN model, whose parameter distribution is summarized in Table~\ref{tab:parameter}, was trained with 40 source-domain utterances and 40 target-domain utterances for 400 epochs to establish a robust mapping between the source $\mathbf{X}^S$ and the target speech $\mathbf{X}^T$.
Through preliminary experiments and sensitivity analysis on the validation set, both $\lambda_{\text{NR}}$ and $\lambda_{\text{CC}}$ in Eq.~\ref{eq_overall} were set to 0.5.
We observed that this configuration provides the optimal trade-off, ensuring that the auxiliary noise and channel reconstruction tasks effectively guide the generation process without overshadowing the primary adversarial objective or degrading the overall generative quality.
Training was performed using the Adam optimizer \cite{kingma2015} with an initial learning rate of 0.0002.
After training, we utilized the generator $G$ and the fine-tuned domain encoders ($B$, $M$) to simulate a dataset. 
Following the setup in \cite{chen2023}, full speech samples from the source domain $S$ and the same 40 target samples from training were used for the generation stage.
Each source sample was transformed into a corresponding target sample using the generator, incorporating dynamic stochastic perturbation to enhance variability.
The generation process produced numerous paired clean and simulated utterances for the training of the downstream ASR and SE models.

From a computational perspective, Table~\ref{tab:parameter} shows that URSA-GAN allocates the majority of parameters to the noise and channel encoders, which are explicitly designed to model domain-specific variations rather than task-specific speech representations.
Unlike recent resource-efficient SE models, such as CTSE-Net, MA-Net, and AFAGC-GinNet \cite{saleem2025,wahab2025,wahab2025a}, which focus on compact codec-based architectures and dual-path time–frequency transformers for real-time enhancement, URSA-GAN is intentionally designed as an offline data simulation framework.
Its objective is not to perform inference-time enhancement, but to generate realistic paired data that improve the robustness of downstream ASR and SE models.
As a result, the increased model capacity primarily affects the data generation stage and does not introduce an additional computational burden during the inference of the downstream systems.
This design choice highlights a fundamental trade-off between model compactness for real-time enhancement and generative expressiveness for cross-domain adaptation, where URSA-GAN prioritizes the latter to maximize cross-task generalization.

\subsection{Downstream Models}

The effectiveness of our framework was assessed through two downstream tasks, namely automatic speech recognition and speech enhancement.
For each task, we selected a lightweight yet high-performing model that supports real-time and edge deployment scenarios.
For ASR, we adopted Whisper\textsubscript{Tiny} \cite{radford2023}, the smallest variant in the Whisper family developed by OpenAI.
The model is based on a multi-layer Transformer architecture and has been pre-trained on 680,000 hours of diverse speech data, covering multiple languages, accents, and background conditions.
Such large-scale and heterogeneous training equips Whisper with strong robustness to a wide range of acoustic and linguistic variations.
Although Whisper\textsubscript{Tiny} has significantly fewer parameters than its larger counterparts, it retains competitive transcription performance, particularly under noisy and mismatched conditions.
Its low computational footprint and efficient inference make it especially suitable for deployment on resource-constrained devices, such as mobile platforms and embedded systems, aligning well with real-time ASR applications in practical environments.
For SE, DEMUCS \cite{defossez2020} was employed, a waveform-domain enhancement model.
DEMUCS features a causal encoder-decoder with U-Net-style skip connections and combines waveform-level and spectrogram-level loss functions for high-fidelity speech reconstruction.
Its ability to handle both stationary and non-stationary noise, as well as reverberation, makes it suitable for real-world noisy scenarios.

\subsection{Evaluation Metrics}

To comprehensively evaluate our framework, we adopted tailored metrics for ASR, SE, and simulated data realism.
For ASR, the character error rate was used, which calculates the ratio of character-level insertions, deletions, and substitutions to the total number of characters, providing an accurate transcription assessment, especially for languages with complex orthographies.
For SE, we employed the perceptual evaluation of speech quality (PESQ) \cite{rix2001} and the short-time objective intelligibility (STOI) \cite{taal2011}.
PESQ estimates the perceptual quality of enhanced speech by comparing it to a clean reference, correlating well with human judgments.
STOI measures speech intelligibility by analyzing temporal and spectral similarity between the enhanced and reference signals, effectively assessing intelligibility under various noise conditions.
To assess simulated data realism, the mean opinion score (MOS) was adopted, where human listeners rate the quality of synthesized speech or audio environments.
The subjective measure offers insights into the naturalness and authenticity of the generated data.

\section{Results and Discussion}

\subsection{Results under Joint Noise and Channel Mismatches}
\label{res:joint}

Table \ref{tab:hat-esc} summarizes model performance on the HAT-ESC dataset, which features both noise and channel mismatches, evaluated on ASR and SE tasks.
The vanilla baseline, which refers to Whisper\textsubscript{Tiny} \cite{radford2023} for ASR and DEMUCS \cite{defossez2020} for SE and was trained solely on source-domain data, shows the most severe degradation under these conditions.
UNA-GAN \cite{chen2023} improves performance by generating target-domain noise through unsupervised adaptation, yielding clear gains over the baseline.
Extending the vanilla model with recordings from multiple devices excluding the target webcam channel (Vanilla + Multi-source) enhances robustness, but the modest gains indicate that source diversity alone cannot fully resolve domain adaptation, and target-domain relevant data remain essential, either labeled or simulated.
Our prior NADA-GAN \cite{wang2024} and CADA-GAN \cite{wang2025} can be viewed as simplified URSA-GAN variants, where NADA-GAN removes the channel encoder and channel consistency loss, and CADA-GAN removes the noise encoder and noise reconstruction loss.
Both incorporate either noise- or channel-aware adaptation with contrastive data simulation, achieving additional gains in ASR and SE.

\begin{table}[t]
\small
\caption{Results on HAT-ESC for downstream ASR (CER, \%) and SE (PESQ) tasks under channel and noise mismatches, with relative improvement (Rel., \%) over the baseline.}
\vspace{-5pt}
\label{tab:hat-esc}
\centering
\setlength{\tabcolsep}{8.5pt}
\begin{tabular}{lcccc}
\toprule
\multirow{2}{*}{\bf{Model}} & \multicolumn{2}{c}{\bf{ASR}} & \multicolumn{2}{c}{\bf{SE}} \\
\cmidrule(lr){2-3}
\cmidrule(lr){4-5}
 & \bf{CER~$\downarrow$} & \bf{Rel.} & \bf{PESQ~$\uparrow$} & \bf{Rel.} \\
\toprule
Vanilla & 32.43 & - & 1.99 & - \\
+ Multi-source & 28.99 & 10.61 & 2.23 & 12.06 \\
UNA-GAN \cite{chen2023} & 29.50 & 9.03 & 2.21 & 11.06 \\
NADA-GAN \cite{wang2024} & 28.77 & 11.29 & 2.25 & 13.07 \\
CADA-GAN \cite{wang2025} & 27.94 & 13.85 & 2.26 & 13.57 \\
\bf{URSA-GAN} & \bf{27.19} & \bf{16.16} & \bf{2.30} & \bf{15.58} \\
+ Multi-source & 28.74 & 11.39 & 2.25 & 13.07 \\
\midrule
Topline & 20.31 & 37.37 & 2.66 & 33.67 \\
\bottomrule
\end{tabular}
\vspace{-10pt}
\end{table}

To address the gap between our experimental setup and realistic development scenarios, where developers often have access to variable-condition source-domain recordings, we additionally included a multi-source variant of URSA-GAN (i.e., URSA-GAN + Multi-source).
More precisely, we treated all seven non-webcam channels as the source domain and evaluated the model under the same setting, where the target domain was defined by the high-variation webcam channel.
Although URSA-GAN + Multi-source does not surpass the single-source URSA-GAN model, this can be attributed to the reduced domain discrepancy between the aggregated multi-source input and the target, which may slightly diminish the generator's incentive to learn strong target-specific transformations.
Nevertheless, URSA-GAN + Multi-source still achieves competitive performance compared to all other baseline systems, even when the source-domain data already cover broad acoustic variability, confirming the robustness and effectiveness of our framework.
The proposed URSA-GAN achieves the best performance, surpassing prior promising GAN-based techniques and the multi-source baseline.
These results confirm the value of jointly modeling noise and channel variations to generate realistic target-domain data, facilitating more robust adaptation for both tasks.

The topline model, trained directly on labeled target-domain data, serves as an upper bound.
In practice, such labeled data are often limited or unavailable, and our evaluation reflects this constraint.
While URSA-GAN does not match the upper bound, the small gap shows its potential to approximate optimal adaptation without large labeled target-domain resources.
Overall, URSA-GAN effectively addresses simultaneous noise and channel mismatches, substantially improving downstream ASR and SE in realistic acoustic environments.

\begin{table}[t]
\small
\caption{CERs (\%) and their relative CER reductions (Rel. \%) on HAT and TAT for the downstream ASR task with channel mismatch.}
\vspace{-5pt}
\label{tab:hat_tat}
\centering
\setlength{\tabcolsep}{6pt}
\begin{tabular}{lcccc}
\toprule
\multirow{2}{*}{\bf{Model}} & \multicolumn{2}{c}{\bf{HAT}} & \multicolumn{2}{c}{\bf{TAT}} \\
\cmidrule(lr){2-3}
\cmidrule(lr){4-5}
 & \bf{CER~$\downarrow$} & \bf{Rel.} & \bf{CER~$\downarrow$} & \bf{Rel.} \\
\toprule
Vanilla \cite{radford2023} & 10.24 & - & 12.76 & - \\
UNA-GAN \cite{chen2023} & 9.76 & 4.69 & 11.82 & 7.37 \\
CADA-GAN \cite{wang2025} & 8.19 & 20.02 & 11.53 & 9.64 \\
\bf{URSA-GAN} & \bf{8.14} & \bf{20.51} & \bf{11.50} & \bf{9.87} \\
\midrule
w/o $\mathcal{L}_{\text{CC}}$ & 8.75 & 14.55 & 11.58 & 9.25 \\
w/o Channel Embeddings & 9.02 & 11.91 & 11.60 & 9.09 \\
\midrule
Topline & 3.88 & 62.11 & 10.30 & 19.28 \\
\bottomrule
\end{tabular}
\vspace{-5pt}
\end{table}

\begin{table}[t]
\small
\caption{CERs (\%) of different Whisper model sizes on HAT and TAT, evaluated before (Raw) and after (Adapted) domain adaptation using URSA-GAN.}
\vspace{-5pt}
\label{tab:whisper_size}
\centering
\setlength{\tabcolsep}{10pt}
\begin{tabular}{lcccc}
\toprule
\multirow{2}{*}{\bf{Model Size}} & \multicolumn{2}{c}{\bf{HAT}} & \multicolumn{2}{c}{\bf{TAT}} \\
\cmidrule(lr){2-3}
\cmidrule(lr){4-5}
 & \bf{Raw} & \bf{Adapted} & \bf{Raw} & \bf{Adapted} \\
\toprule
Tiny & 10.24 & \bf{8.14} & 12.76 & \bf{11.50} \\
Base & 7.62 & \bf{6.35} & 11.42 & \bf{11.03} \\
Small & 4.51 & \bf{4.05} & 10.65 & \bf{10.57} \\
Medium & 3.88 & \bf{3.25} & 8.15 & \bf{7.56} \\
\bottomrule
\end{tabular}
\vspace{-10pt}
\end{table}

\subsection{Speech Recognition under Channel Mismatch}

Table \ref{tab:hat_tat} presents CER results of various domain adaption methods on HAT and TAT. 
URSA-GAN outperforms the vanilla baseline (Whisper\textsubscript{Tiny}) \cite{radford2023} and also improves upon our previously proposed CADA-GAN \cite{wang2025}, achieving substantial relative CER reductions on both corpora. 
The slight advantage over CADA-GAN can be attributed to the inclusion of the noise encoder, which complements the channel encoder by capturing acoustic information that the channel encoder may overlook.
These results highlight the effectiveness of incorporating pre-trained encoders to mitigate channel mismatch.
Notably, although the channel encoder was trained solely on HAT, the improvements extend to TAT, indicating that the encoder captures transferable channel-specific features independent of phonetic content.
This generalization is critical for robust ASR across diverse linguistic and channel conditions.

The contributions of individual components were further examined.
Removing the channel consistency loss slightly degrades performance, suggesting it aids channel fidelity but has limited direct impact on ASR accuracy.
Omitting channel embeddings, however, leads to a substantial performance drop, emphasizing their central role in transferring essential target-domain channel characteristics and enhancing model robustness across varying channels.

\begin{table}[t]
\small
\caption{PESQ, average ranks (Avg. Rank), and STOI (\%) on VBD for the downstream SE task under noise mismatch, with average ranks derived from PESQ scores across all test samples using the Friedman test.}
\vspace{-5pt}
\label{tab:vbd}
\centering
\setlength{\tabcolsep}{1.5pt}
\begin{tabular}{lccc}
\toprule
\bf{Model} & \bf{PESQ~$\uparrow$} & \bf{Avg. Rank~$\downarrow$} & \bf{STOI~$\uparrow$} \\
\toprule
Noisy & 2.20 & - & 88.9 \\
Vanilla \cite{defossez2020} & 3.05 & 4.94 & \bf{95.2} \\
RemixIT \cite{tzinis2022} & 3.07 & 4.42 & 95.1 \\
UNA-GAN \cite{chen2023} & 3.09 & 3.89 & 95.0 \\
NADA-GAN \cite{wang2024} & 3.14 & 2.36 & 95.2 \\
URSA-GAN (Pre-trained BEATs) & 3.10 & 3.52 & 95.0 \\
\bf{URSA-GAN (Fine-tuned BEATs)} & \bf{3.16} & \bf{1.88} & \bf{95.3} \\
\midrule
w/o Perturbation & 3.13 & - & 95.2 \\
w/o $\mathcal{L}_{\text{NR}}$ & 3.12 & - & 95.2 \\
w/o Noise Embeddings & 3.11 & - & 95.2 \\
\midrule
Topline & 3.11 & - & 95.1 \\
\bottomrule
\end{tabular}
\vspace{-10pt}
\end{table}

\subsection{Effect of Domain Adaptation Across Whisper Model Sizes}

Table \ref{tab:whisper_size} reports the CERs of different Whisper \cite{radford2023} model sizes on the HAT and TAT corpora, evaluated before and after domain adaptation using  our URSA-GAN.
Across all model sizes, domain adaptation consistently reduces CER, demonstrating the effectiveness of our framework in mitigating both noise and channel mismatches.
Smaller models such as Whisper\textsubscript{Tiny} and Whisper\textsubscript{Base} exhibit larger absolute reductions, reflecting the greater sensitivity of lightweight models to domain shifts.
As model size increases, the raw CER naturally decreases, indicating stronger intrinsic generalization capabilities of larger Whisper variants.
Adaptation further refines performance, yielding relative gains across models, although the absolute improvement diminishes due to the already low error rates.
These results highlight that URSA-GAN provides consistent benefits across different ASR model scales, enhancing robustness in low-resource and realistic cross-domain scenarios.
Note that we did not evaluate Whisper\textsubscript{Large} because of hardware resource limitations, but the observed trend suggests that combining domain adaptation with larger pre-trained models can achieve near-optimal recognition performance while maintaining flexibility for deployment in diverse acoustic conditions.

\subsection{Speech Enhancement under Noise Mismatch}

Table \ref{tab:vbd} presents PESQ and STOI results on VBD.
URSA-GAN consistently outperforms the vanilla baseline (DEMUCS) \cite{defossez2020} and improves upon our earlier NADA-GAN \cite{wang2024}.
The additional channel encoder further captures acoustic cues that the noise encoder alone cannot.
A pre-trained BEATs model without fine-tuning yields only modest gains, underscoring the need for target-domain adaptation.
URSA-GAN also surpasses models trained solely on limited real noisy data (Topline), showing that structured noise simulation generalizes better than naive augmentation.
We additionally compared against a RemixIT-style strategy \cite{tzinis2022}, which augments clean speech with predicted target noise.
While this strategy improves over the vanilla baseline, it underperforms GAN-based methods as it cannot capture complex speech–noise interactions or temporal dynamics, and with limited unlabeled target data it risks overfitting to specific noise types.
In contrast, UNA-GAN and URSA-GAN generate diverse noisy examples via a learned clean-to-noisy transformation.
To confirm statistical significance, we conducted a Friedman test \cite{friedman1937} on PESQ scores across 784 test utterances.
The result ($\chi^2(5)=1568.35$, $p<0.0001$) indicates strong system-level differences.
Post-hoc Nemenyi tests \cite{demsar2006} further show that URSA-GAN with the fine-tuned BEATs achieves the best average rank (1.88), significantly outperforming the vanilla baseline, RemixIT, UNA-GAN, URSA-GAN with the pre-trained BEATs, and NADA-GAN ($p<0.05$).
Although its margin over NADA-GAN is small, URSA-GAN consistently ranks highest.
Ablation studies indicate that removing dynamic stochastic perturbation slightly reduces PESQ, omitting the noise reconstruction loss has a larger impact, and discarding noise embeddings causes the most severe degradation, confirming their critical role in transferring target-domain noise characteristics.

\begin{figure}
\centering
\includegraphics[width=1.0\linewidth]{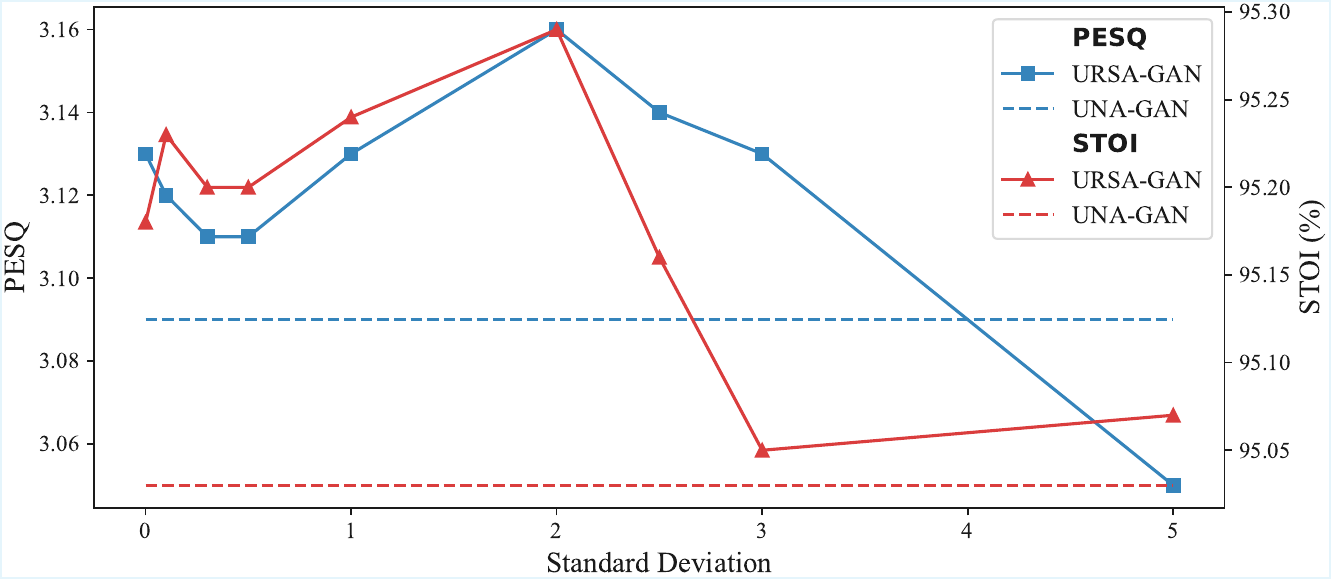}
\vspace{-20pt}
\caption{PESQ and STOI results of dynamic stochastic perturbation under various standard deviations on the VBD dataset.}
\label{fig:dsp}
\vspace{-5pt}
\end{figure}

Fig. \ref{fig:dsp} further reveals that moderate perturbation prevents overfitting and maximizes PESQ, whereas excessive perturbation degrades quality.
STOI remains stable across perturbation levels, indicating intelligibility is less sensitive than perceptual quality.
These results highlight the importance of balancing noise simulation variability for optimal enhancement.

\begin{table}[t]
\small
\caption{Comparison of various SE models on VBD, reporting PESQ and STOI (\%) before (Raw) and after (Adapted) domain adaptation using URSA-GAN.}
\vspace{-5pt}
\label{tab:se}
\centering
\setlength{\tabcolsep}{9pt}
\begin{tabular}{lcccc}
\toprule
\multirow{2}{*}{\bf{Model}} & \multicolumn{2}{c}{\bf{PESQ~$\uparrow$}} & \multicolumn{2}{c}{\bf{STOI~$\uparrow$}} \\
\cmidrule(lr){2-3}
\cmidrule(lr){4-5}
 & \bf{Raw} & \bf{Adapted} & \bf{Raw} & \bf{Adapted} \\
\toprule
Noisy & 2.20 & - & 88.9 & - \\
DEMUCS \cite{defossez2020} & 3.05 & \bf{3.16} & 95.2 & \bf{95.3} \\
MP-SENet \cite{lu2023} & 3.53 & \bf{3.59} & 96.1 & \bf{96.3} \\
SEMamba \cite{chao2024} & 3.71 & \bf{3.74} & 96.2 & \bf{96.3} \\
\bottomrule
\end{tabular}
\vspace{-10pt}
\end{table}

\subsection{Adaptation Benefits Across Various Enhancement Models}

Table \ref{tab:se} details the performance of various SE models on the VBD dataset, evaluated in terms of PESQ and STOI before and after domain adaptation using URSA-GAN.
All models benefit from adaptation, with improvements in both perceptual quality and intelligibility metrics.
Notably, models with lower baseline performance, such as DEMUCS \cite{defossez2020}, exhibit more substantial gains, while stronger models like MP-SENet \cite{lu2023} and SEMamba \cite{chao2024} show smaller yet consistent improvements.
This trend aligns with the expectation that domain adaptation has a larger effect when the original model is less robust to target-domain mismatches.
These results demonstrate that URSA-GAN effectively enhances a wide range of SE architectures by bridging the gap between source- and target-domain conditions, confirming the generalizability of our adaptation framework.

\begin{figure}[t]
\centering
\includegraphics[width=1.0\linewidth]{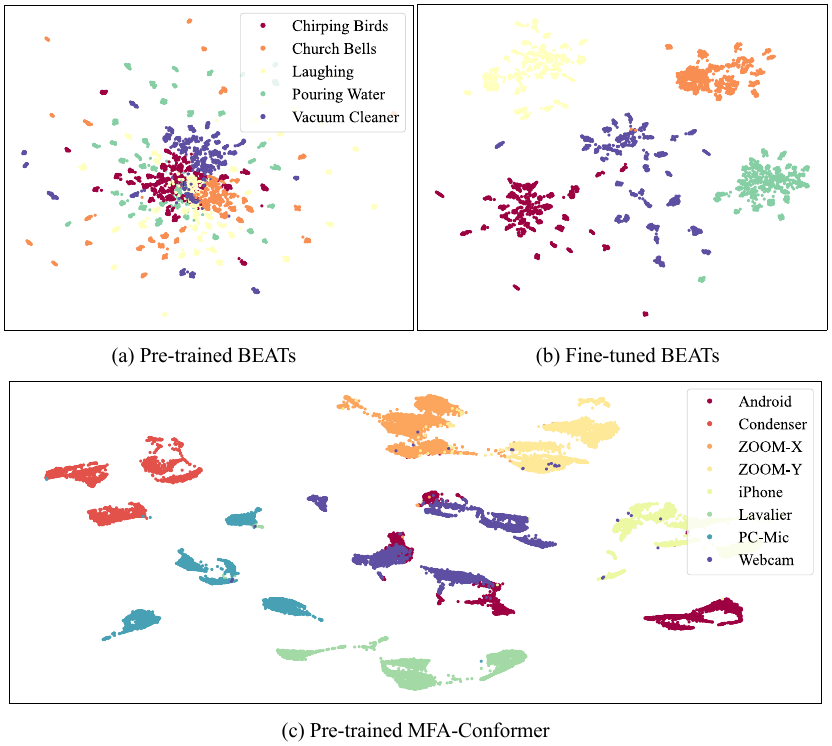}
\vspace{-20pt}
\caption{The UMAP visualization of embeddings extracted from different noise and channel types in the HAT-ESC dataset.}
\label{fig:umap}
\vspace{-10pt}
\end{figure}

\subsection{Visualization of Embeddings}

To analyze the learned noise representations, Uniform Manifold Approximation and Projection (UMAP) \cite{mclnnes2018} was applied to visualize the noise embeddings extracted from the five unseen noise types in the HAT-ESC test set.
Fig. \ref{fig:umap} (a) and (b) illustrate the embeddings obtained from the pre-trained BEATs and fine-tuned BEATs models \cite{chen2023a}, respectively.
Compared to the pre-trained embeddings, the fine-tuned BEATs embeddings exhibit a more distinct separation between different noise categories, including non-stationary noise types.
The improved discriminability underscores the effectiveness of fine-tuning in adapting the noise encoder to the target domain, thereby enhancing the ability of our model to differentiate between diverse noise profiles.
This capability plays a crucial role in the improved performance of our proposed framework.

Furthermore, to gain deeper insights into the behavior of URSA-GAN, we visualized the learned channel embeddings in Fig. \ref{fig:umap} (c).
The results reveal a clear separation among the eight different channel types, even though the MFA-Conformer \cite{zhang2022} was trained solely on six channels in the HAT-ESC dataset, excluding both the source and target channels.
This demonstrates that our pre-trained channel encoder effectively captures the unique acoustic characteristics associated with each microphone.
The distinct clustering further validates the robustness of our framework in modeling channel variations, highlighting its generalization capability to unseen channels.

\subsection{Impact of Feature Fusion Strategies and FiLM Variants}

Table \ref{tab:fusion} compares different feature fusion strategies and FiLM \cite{perez2018} variants on the HAT-ESC dataset for ASR and SE.
Naive fusion approaches, including concatenation and addition of target embeddings, produce only modest gains, which suggests that simple combination of auxiliary features is insufficient for effective adaptation.
Introducing FiLM at the encoder layer already improves performance and demonstrates that conditional feature modulation helps the generator adjust to varying noise and channel conditions.
Extending FiLM to all layers except the decoder layer with shared parameters further enhances results and shows the benefit of deeper integration.
Assigning independent FiLM parameters to the encoder layer and each ResNet block achieves the largest improvements and indicates that block-specific modulation is crucial for fully exploiting noise and channel embeddings.
These findings highlight the importance of hierarchical and independent feature modulation, as progressively incorporating FiLM across layers and blocks allows the model to adapt more precisely to target-domain conditions.

\begin{table}[t]
\small
\caption{Impact of different feature fusion strategies on ASR (CER, \%) and SE (PESQ) performance using the HAT-ESC dataset.}
\vspace{-5pt}
\label{tab:fusion}
\centering
\setlength{\tabcolsep}{8pt}
\begin{tabular}{lcc}
\toprule 
\bf{Fusion Strategy} & \bf{CER~$\downarrow$} & \bf{PESQ~$\uparrow$} \\
\toprule
Concatenation & 28.23 & 2.24 \\
Addition & 28.11 & 2.25 \\
\midrule
\multicolumn{3}{l}{\textit{\textbf{FiLM}}} \\
Encoder Only & 27.84 & 2.27 \\
All Layers, Shared Parameters & 27.55 & 2.28 \\
\bf{All Layers, Independent Parameters} & \bf{27.19} & \bf{2.30} \\
\bottomrule
\end{tabular}
\vspace{-5pt}
\end{table}

\begin{table}[t]
\small
\caption{Downstream ASR (CER, \%) and SE (PESQ) performance of URSA-GAN with different noise encoders on HAT-ESC, including encoder parameter counts for size comparison.}
\vspace{-5pt}
\label{tab:noise_encoder}
\centering
\setlength{\tabcolsep}{9pt}
\begin{tabular}{lccc}
\toprule 
\bf{Noise Encoder} & \bf{CER~$\downarrow$} & \bf{PESQ~$\uparrow$} & \bf{Parameters (M)} \\
\toprule
\multicolumn{4}{l}{\textit{\textbf{Whisper}} \cite{radford2023}} \\
Tiny & 28.53 & 2.20 & \bf{39.0} \\
Base & 28.12 & 2.22 & 74.0 \\
Small & 27.54 & 2.26 & 244.0 \\
\midrule
\multicolumn{4}{l}{\textit{\textbf{WavLM}} \cite{chen2022c}} \\
Base & 27.84 & 2.25 & 90.7 \\
Large & 27.28 & 2.29 & 316.6 \\
\midrule
\textbf{BEATs} \cite{chen2023a} & \bf{27.19} & \bf{2.30} & 90.0 \\
\bottomrule
\end{tabular}
\vspace{-10pt}
\end{table}

\subsection{Evaluation of Noise Encoder Selection}

Table~\ref{tab:noise_encoder} presents the performance of URSA-GAN when adopting different noise encoders to provide noise embeddings for downstream ASR and SE tasks on HAT-ESC.
We considered Whisper\textsubscript{Tiny} \cite{radford2023}, Whisper\textsubscript{Base}, Whisper\textsubscript{Small}, WavLM\textsubscript{Base} \cite{chen2022c}, WavLM\textsubscript{Large}, and BEATs \cite{chen2023a}, reporting CER for ASR and PESQ for SE, along with parameter counts for size comparison of encoders.
The results indicate that noise encoders originally trained for environmental sound recognition yield more informative representations for adaptation.
Among the tested encoders, BEATs achieves the lowest CER and the highest PESQ, confirming the advantage of leveraging pre-training on diverse non-speech acoustic events.
WavLM provides moderate improvements due to its speech-oriented but noise-robust features, while Whisper variants, primarily designed for ASR, show the least benefit when repurposed for noise representation.
These findings highlight the critical role of encoder selection, since specialized environmental sound encoders such as BEATs are more effective at capturing target-domain noise statistics and enable the generator to synthesize speech that reflects realistic acoustic conditions while preserving intelligibility.

\begin{figure}[t]
\centering
\includegraphics[width=1.0\linewidth]{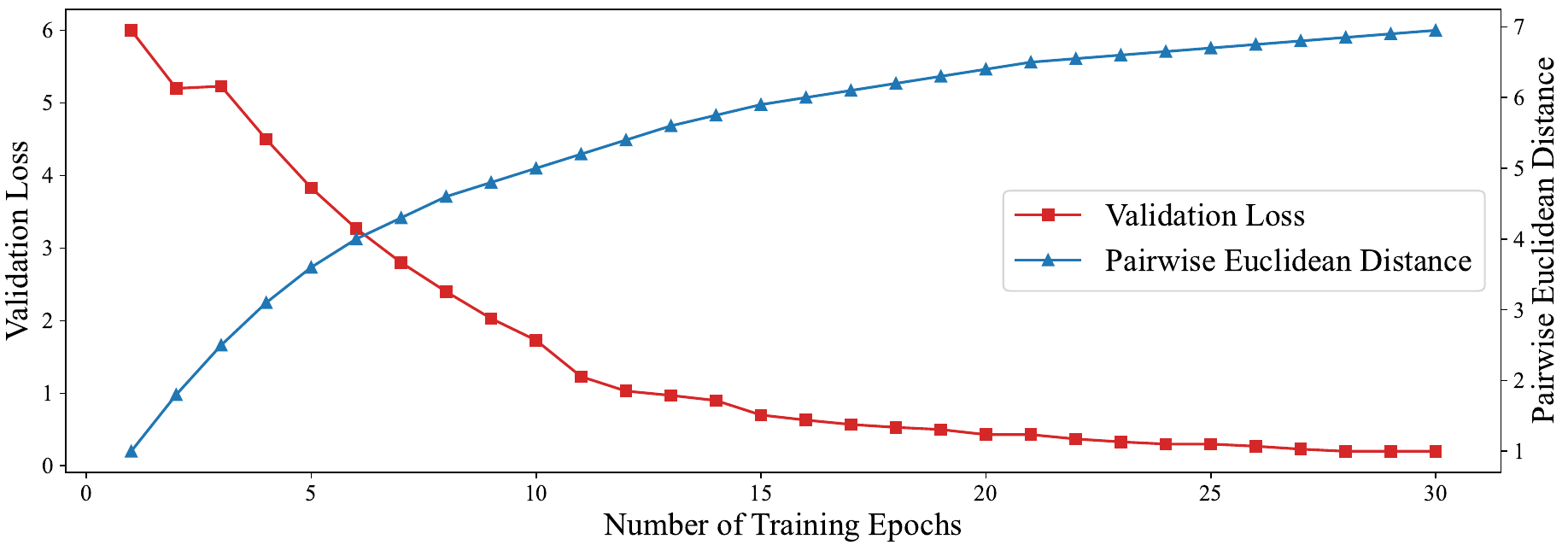}
\vspace{-20pt}
\caption{Validation loss of our channel encoder on the HAT-ESC dataset and the average pairwise Euclidean distance between channel embeddings across training epochs.}
\label{fig:distance}
\vspace{-5pt}
\end{figure}

\begin{table}[t]
\small
\caption{MOSs of simulated speech generated by our framework and the baseline method across various datasets.}
\vspace{-5pt}
\label{tab:mos}
\centering
\setlength{\tabcolsep}{6.5pt}
\begin{tabular}{lccc}
\toprule 
\bf{Model} & \bf{HAT} & \bf{TAT} & \bf{VBD} \\
\toprule
UNA-GAN \cite{chen2023} & 2.90 $\pm$ 0.75 & 2.55 $\pm$ 1.11 & 1.49 $\pm$ 0.42\\
\bf{URSA-GAN} & \bf{4.06 $\pm$ 0.71} & \bf{3.09 $\pm$ 1.06} & \bf{2.51 $\pm$ 0.62}\\
\bottomrule
\end{tabular}
\vspace{-10pt}
\end{table}

\subsection{Analysis of Embedding Divergence and Validation Loss}

To quantitatively analyze the channel discrimination ability of our channel encoder, we tracked the training progress by monitoring both the validation loss and the average pairwise Euclidean distance $d$ between channel embeddings.
The pairwise distance, computed using the validation set of the HAT-ESC dataset, was averaged between embeddings of the same utterance set, indicating the same speaker and content recorded across different channels.
The metric is given by
\begin{equation}
d = \frac{\sum_{u \in \mathcal{U}} \sum_{1 \leq i < j \leq 8} \|\mathbf{c}_i^u - \mathbf{c}_j^u\|_2}{|\mathcal{U}| \cdot \binom{8}{2}},
\end{equation}
where $\mathbf{c}_i^u$ and $\mathbf{c}_j^u$ represent the channel embeddings for channels $i$ and $j$ from the same utterance set $u$, and $\mathcal{U}$ denotes the set of utterances used in validation.
As shown in Fig. \ref{fig:distance}, the validation loss (red curve) steadily decreases over 30 epochs, while the pairwise distance (blue curve) increases.
This indicates that the channel encoder is effectively learning to distinguish between different channel characteristics over the course of training.
The increasing distance between embeddings suggests that our channel encoder is successfully separating channel-specific information, which is critical for improving model performance.
The positive trend underscores the essential role of the channel encoder in the success of the URSA-GAN framework.

\subsection{MOS Assessment on Simulated Data}

To assess the perceptual realism of our generated speech, we conducted a comprehensive MOS assessment focusing on noise and channel similarity with real target-domain recordings.
Ten participants rated the similarity between generated and real audio on a scale of 1 to 5.
As shown in Table \ref{tab:mos}, URSA-GAN outperforms the UNA-GAN baseline \cite{chen2023}, demonstrating the ability of our model to better replicate the acoustic characteristics of the target domain.
Despite not being explicitly trained on the TAT corpus, URSA-GAN achieves higher MOS scores, highlighting its strong generalization.
Furthermore, URSA-GAN exhibits a lower standard deviation, indicating a more consistent perceptual quality.
These results confirm its effectiveness in generating realistic speech, even with limited target-domain data, making it valuable for speech processing applications.

\begin{table}[t]
\small
\caption{SE effectiveness (PESQ) and ASR performance (CER, \%) of URSA-GAN across varying SNR conditions on HAT-ESC.}
\vspace{-5pt}
\label{tab:snr}
\centering
\setlength{\tabcolsep}{4.5pt}
\begin{tabular}{lcccccc}
\toprule
\multirow{2}{*}{\bf{SNR}} & \multicolumn{3}{c}{\bf{PESQ~$\uparrow$}} & \multicolumn{3}{c}{\bf{CER~$\downarrow$}} \\
\cmidrule(lr){2-4}
\cmidrule(lr){5-7}
 & \bf{Raw} & \bf{Adapted} & \bf{Gain} & \bf{Raw} & \bf{Adapted} & \bf{Error Type} \\
\toprule
0 & 1.05 & 2.10 & 1.05 & 35.83 & 28.51 & Substitution \\
5 & 1.41 & 2.37 & 0.96 & 28.09 & 22.10 & Deletion \\
10 & 1.77 & 2.51 & 0.74 & 20.53 & 21.01 & Insertion \\
15 & 2.09 & 2.56 & 0.47 & 14.98 & 14.54 & Substitution \\
\bottomrule
\end{tabular}
\vspace{-10pt}
\end{table}

\subsection{Performance Across Various SNR Levels}

The proposed framework demonstrates strong effectiveness in handling domain mismatches, as discussed in Section \ref{res:joint}. 
To further characterize its behavior, we evaluated performance across various SNR conditions on the HAT-ESC dataset in Table \ref{tab:snr}. 
URSA-GAN consistently reduces CER and improves PESQ under low-SNR conditions, indicating robustness in noisy environments. 
At very low SNRs, substitution and deletion errors are more prevalent, reflecting challenges in modeling fine acoustic cues and transient energy. 
At moderate SNRs, mild insertion errors may occur due to subtle artifacts in the generated speech, slightly affecting CER despite PESQ improvements. 
At higher SNRs, both CER and PESQ stabilize, though minor distortions can still induce occasional substitution errors. 
These observations suggest that URSA-GAN effectively generates target-domain speech across noise conditions, and further incorporation of ASR-sensitive feedback could enhance downstream performance.

\section{Conclusion and Future Work}

This study\footnote{Code: \url{https://github.com/JethroWangSir/URSA-GAN/}.} puts forward URSA-GAN, a unified generative framework that addresses the challenge of adapting speech recognition and enhancement models under joint noise and channel mismatches.
By integrating dual pre-trained encoders for noise and channel characteristics into a GAN-based architecture, URSA-GAN generates speech that is acoustically consistent with the target domain while preserving phonetic integrity.
The use of dynamic stochastic perturbation further enhances the generalization of the models, especially when only limited target-domain data are available.
Extensive experiments across multiple datasets and tasks confirm the effectiveness of our framework.
URSA-GAN consistently outperforms strong baselines in both ASR and SE settings.
Notably, it even surpasses models trained directly on real target data in certain conditions, highlighting the benefits of structured domain simulation.
Ablation studies confirm the critical role of both noise and channel components, with channel modeling contributing slightly more to performance gains.
Despite these promising results, several limitations remain.
First, like many GAN-based architectures, URSA-GAN requires careful hyperparameter tuning to ensure training stability and prevent mode collapse.
Second, the reliance on multiple large-scale pre-trained encoders (BEATs and MFA-Conformer) introduces additional computational overhead during the training phase, making the data simulation process resource-intensive compared to simple augmentation methods.
However, this overhead is incurred only once during offline training and does not affect the inference efficiency of the downstream models.

Future directions include improving the adaptability of the pre-trained encoders through domain-aware or self-supervised pre-training schemes.
Another promising direction is to explore alternative generative backbones, such as diffusion models, to further enhance the quality of the synthesis.
Additionally, extending URSA-GAN to support more diverse and temporally dynamic noise and channel conditions would make it more applicable to real-world deployments.
Finally, integrating the framework more tightly with end-to-end ASR and SE pipelines could enable seamless adaptation and jointly optimized performance across tasks.

\bibliographystyle{IEEEtran}
\bibliography{references.bib}

\begin{IEEEbiography}
[{\includegraphics[width=1in,height=1.25in,clip,keepaspectratio]{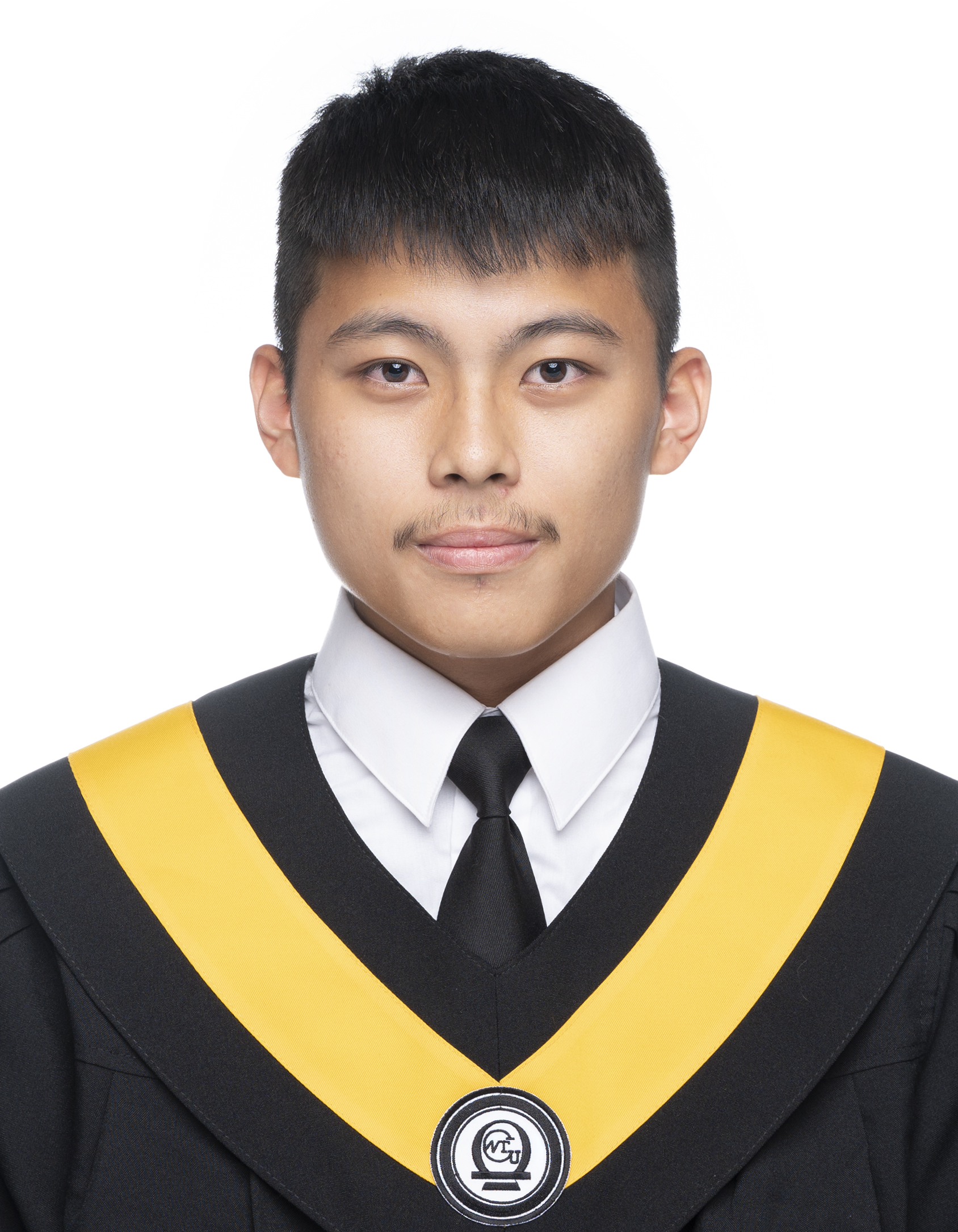}}]
{Chien-Chun Wang}(Student Member, IEEE) received the M.S. degree in computer science and information engineering from National Taiwan Normal University (NTNU), Taipei, Taiwan, in 2025. He currently serves as an Engineer with E.SUN Financial Holding Co., Ltd., Taipei, Taiwan. His research focuses on speech recognition, speech enhancement, voice activity detection, and speech quality assessment.
\end{IEEEbiography}

\begin{IEEEbiography}
[{\includegraphics[width=1in,height=1.25in,clip,keepaspectratio] {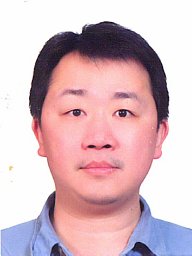}}]
{Hung-Shin Lee} received the Ph.D. degree in electrical engineering from National Taiwan University (NTU), Taipei, Taiwan, in 2021. He is currently an AI Engineer with United Link Co., Ltd., Taiwan. Since February 2026, he has also been an Adjunct Assistant Professor with Graduate Institute of AI Interdisciplinary Applied Technology, National Taiwan Normal University, Taipei, Taiwan. His research interests include speech recognition and synthesis, as well as speech and video enhancement.
\end{IEEEbiography}

\begin{IEEEbiography}
[{\includegraphics[width=1in,height=1.25in,clip,keepaspectratio] {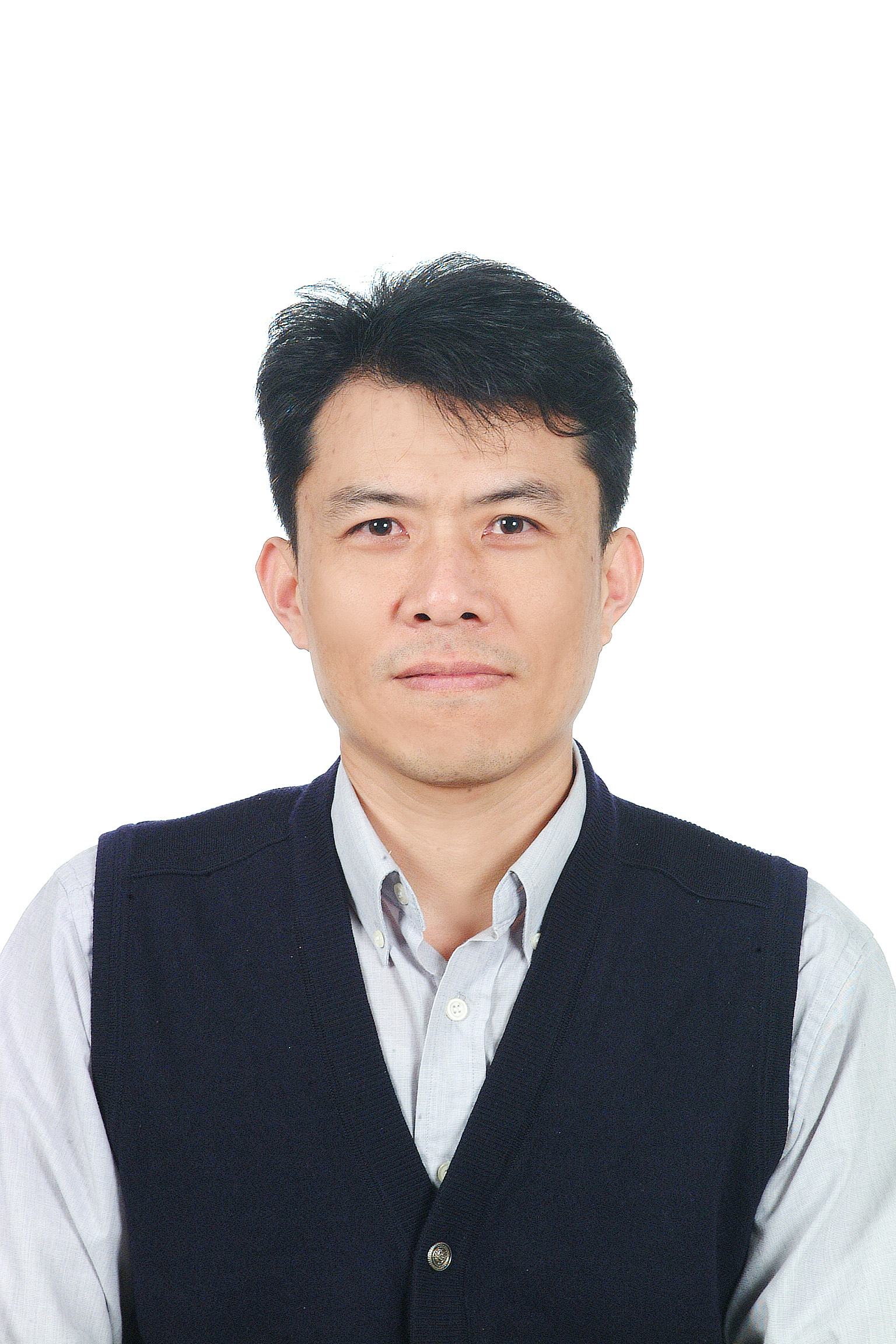}}]
{Hsin-Min Wang}(Senior Member, IEEE) received the B.S. and Ph.D. degrees in electrical engineering from National Taiwan University, Taipei, Taiwan, in 1989 and 1995, respectively. In October 1995, he joined the Institute of Information Science, Academia Sinica, Taipei, Taiwan, where he is currently a Research Fellow. He was a Joint Professor in the Department of Computer Science and Information Engineering at National Cheng Kung University from 2014 to 2023. He was an Associate Editor of IEEE/ACM Transactions on Audio, Speech, and Language Processing from 2016 to 2020. He currently serves a Senior Editor of APSIPA Transactions on Signal and Information Processing. His major research interests include spoken language processing, natural language processing, multimedia information retrieval, and machine learning. He was a General Co-Chair of ISCSLP2016, ISCSLP2018, ASRU2023, and O-COCOSDA2024 and a Technical Co-Chair of ISCSLP2010, O-COCOSDA2011, APSIPAASC2013, ISMIR2014, ASRU2019, and APSIPAASC2023. He received the Chinese Institute of Engineers Technical Paper Award in 1995 and the ACM Multimedia Grand Challenge First Prize in 2012. He was an APSIPA distinguished lecturer for 2014–2015. He is a member of ISCA, ACM, and APSIPA.
\end{IEEEbiography}

\begin{IEEEbiography}
[{\includegraphics[width=1in,height=1.25in,clip,keepaspectratio] {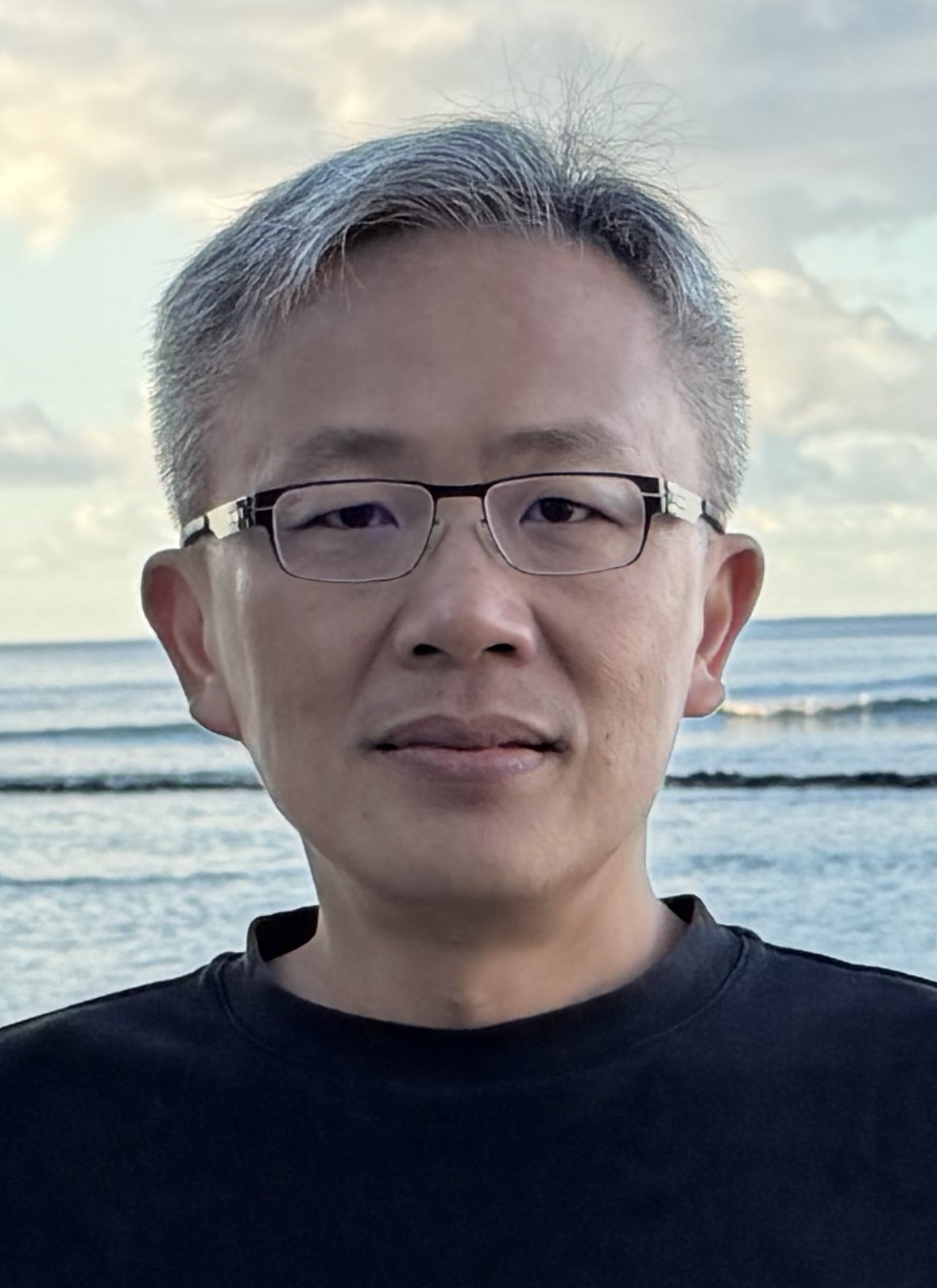}}]
{Berlin Chen}(Member, IEEE) is a Distinguished Professor of the Computer Science and Information Engineering Department at National Taiwan Normal University (NTNU), Taipei, Taiwan. He received his Ph.D. degree in computer science and information engineering from National Taiwan University (NTU) in 2001, and then joined NTNU as an Assistant Professor in 2002. He became a Professor in 2010. Prof. Chen's research interests generally lie in the areas of speech recognition and natural language processing, multimedia information retrieval, computer-assisted language learning, and machine learning in general.
\end{IEEEbiography}

\end{document}